\documentclass[journal]{IEEEtran}

\usepackage[numbers,sort&compress]{natbib}
\usepackage{amsmath}
\usepackage{stmaryrd}
\usepackage{amssymb}
\usepackage{graphicx}
\usepackage{epstopdf}
\usepackage{stfloats}
\usepackage{dsfont}

\graphicspath{{fig/}}
\DeclareGraphicsExtensions{.eps}
\usepackage{epstopdf}
\usepackage{algorithm}
\usepackage{algorithmic}

\begin{document}
%
\title{A Hybrid ARQ Scheme Based on Polar Codes}
%
%
%
\author{Kai~Chen,~\IEEEmembership{Student Member,~IEEE,}
        Kai~Niu,~\IEEEmembership{Member,~IEEE,}
        and~Jia-Ru~Lin,~\IEEEmembership{Member,~IEEE}
\thanks{This work was supported in part by the National Basic Research Program of China (973 Program) (No. 2009CB320401), the National Natural Science Foundation of China (No. 61171099), the National Science and Technology Major Project of China (No. 2012ZX03003-007) and Qualcomm Corporation.}
\thanks{The authors are with the Key Laboratory of Universal Wireless Communication, Ministry of Education,
Beijing University of Posts and Telecommunications,
Beijing 100876, China. (e-mail: \{kaichen, niukai, jrlin\}@bupt.edu.cn)}
}

%

\maketitle

\newtheorem{theorem}{Theorem}
\newtheorem{example}{Example}

\begin{abstract}

A hybrid automatic repeat request (HARQ) scheme based on a novel class of rate-compatible polar (\mbox{RCP}) codes are proposed.
The RCP codes are constructed by performing punctures and repetitions on the conventional polar codes.
Simulation results over binary-input additive white Gaussian noise channels (BAWGNCs) show that, using a low-complexity successive cancellation (SC) decoder, the proposed HARQ scheme performs as well as the existing schemes based on turbo codes and low-density parity-check (LDPC) codes.
The proposed transmission scheme is only about 1.0-1.5dB away from the channel capacity with the information block length of 1024 bits.

\end{abstract}

\begin{IEEEkeywords}
Polar codes, hybrid ARQ, rate-compatible coding, successive cancellation decoding.
\end{IEEEkeywords}

\section{Introduction}
\IEEEPARstart{P}{olar} codes, recently introduced by Ar{\i}kan \cite{Arikan_first}, are the first structured codes that provably achieve the symmetric capacity of binary-input memoryless channels (BMCs).
Given a BMC $W$, by performing the channel polarizing transformation over $N$ independent copies of $W$, we get a second set of synthesized BMCs.
As the transformation size $N$ goes infinity, some of the resulting channels tend to completely noised, and the others tend to noise-free, where the fraction of the latter approaches the symmetric capacity of $W$.
By transmitting free bits over the noiseless channels and sending fixed bits over the others, polar codes can achieve the symmetric capacity under a successive cancellation (SC) decoder with both encoding and decoding complexity $O\left( N\log N \right)$.

\begin{figure}[!t]
  \centering
  \includegraphics[width=0.9\columnwidth]{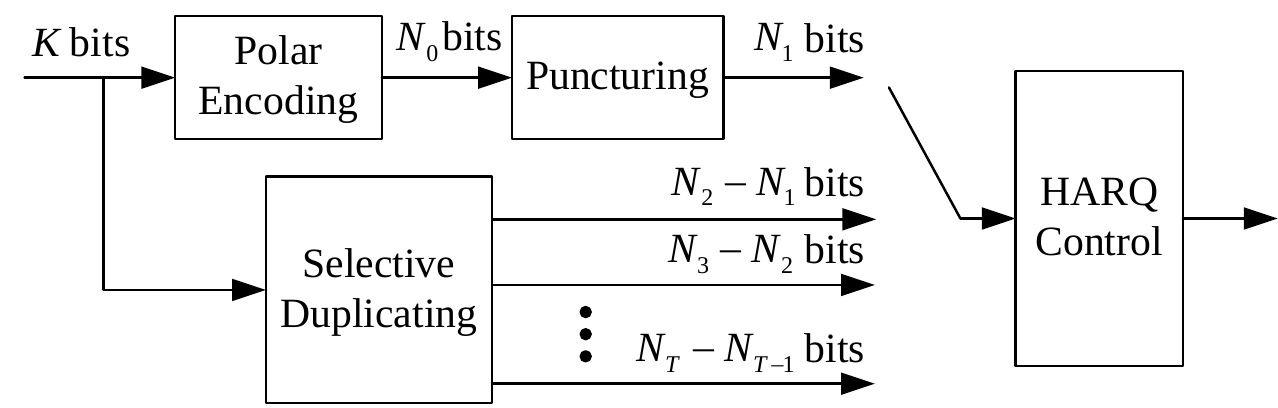}
  \caption{Transmitter block diagram of the proposed HARQ scheme.}
  \label{fig_diagram}
\end{figure}

In delay insensitive communications, hybrid automatic repeat request (HARQ) is widely used to obtain capacity-approaching throughput efficiency \cite{Hagenauer_RCPC}, \cite{Rowitch_RCPT}, \cite{Yue_RCIRA}.
This paper focuses on designing an HARQ transmission scheme of polar codes by using punctures and repetitions.
The proposed transmitter block diagram is shown in Fig. \ref{fig_diagram}.
As the initial transmission, an information block of $K$ bits is fed into a polar encoder. The output codeword of $N_0$ bits are punctured into $N_1$ bits and sent over the channel.
If the receiver fails to decode the codeword, an NACK (negative acknowledgement) is sent to the transmitter through the feedback channel.
And then, $N_2-N_1$ of the information bits are retransmitted.
The receiver tries to perform decoding with all the received $N_2$ bits.
If the decoding is failed again, another $N_3-N_2$ bits are transmitted.
This process continues until the transmitter receives an ACK (acknowledgement), or a maximum number of transmissions $T$ is achieved.

The remainder of the paper is organized as follows. \mbox{Section \ref{section_RCPrP}} describes the rate-compatible polar (RCP) codes used in the proposed HARQ scheme.
Section \ref{section_harq} proposes the design method of the HARQ scheme based on RCP codes.
Section \ref{section_simulation} provides the simulation comparison of HARQ schemes based on polar codes, turbo codes and LDPC codes.
Finally, \mbox{Section \ref{section_conclusions}} concludes the paper.

\section{Rate-Compatible Polar Codes}
\label{section_RCPrP}
In this paper, we use calligraphic characters, such as $\mathcal{X}$ to denote a set.
Let $|\mathcal{X}|$ denote the number of elements in $\mathcal{X}$.
We write lowercase letters (e.g., $x$) to denote scalars, bold-face lowercase letters (e.g., $\textbf{x}$) to denote vectors, and $x_i$ to denote the $i$-th element of $\textbf{x}$.
For any $i \le j$, $\textbf{x}_{i:j}$ denotes a subvector of $\textbf{x}$, i.e., $\textbf{x}_{i:j}=({x}_{i}, {x}_{i+1}, \cdots, {x}_{j})$.
The bold numbers $\textbf{0}$ and $\textbf{1}$ are used to denote the all-zero and all-one vectors, respectively.
Throughout this paper, the base of the logarithm is $2$.

The proposed transmission scheme depicted in Fig. \ref{fig_diagram} is based on a class of RCP codes with punctures and repetitions.
The coded bits of a RCP code consist of two parts: the first part come from a punctured polar encoder, and these bits are called \emph{polar bits}; the other bits are the selective copies of the information bits and are called \emph{repetition bits}.
Hereafter, the RCP code is represented by a three-tuple $(N, K, M)$, $K \le M \le N$ , which means this code has an information block length $K$, and the resulting $N$-length codeword consists of $M$ polar bits and $N-M$ repetition bits.

To construct a $(N, K, M)$ RCP code over a BMC $W$, we first need to construct a punctured polar code with code length $M$.
The punctured polar coding is discussed in \cite{Niu_icc}.

Similar to constructing a conventional polar code, after performing a polarization transform on $N_0=2^{\lceil \log M \rceil}$ independent uses of $W$, we get ${N_0}$ successive uses of synthesized binary input channels $W_{N_0}^{(i)}$, $i=1,2,\cdots,N_0$, where $\lceil \cdot \rceil$ is the top integral function.

Given a symmetric BMC $W$, let $\textbf{a}$ denotes the probability density function (pdf) of the log-likelihood ratio (LLR) of the received bit when a bit zero is transmitted.
The reliability of $W$ can be measured as the error probability
\begin{equation}
\label{equ_bitpe}
P_e(W) = \int \nolimits_{-\infty}^{0} \textbf{a}(x) \text{d}x
\end{equation}
Let $\textbf{a}_{N_0}^{(i)}$, $i=1,2,\cdots,N_0$ denote the LLR pdfs of the received bit from $W_{N_0}^{(i)}$ when all-zero information bits are transmitted.
After calculating $\textbf{a}_{N_0}^{(i)}$ by density evolution (DE) \cite{Mori_DE}, the reliabilities of $W_{N_0}^{(i)}$ are determined by (\ref{equ_bitpe}).
In transmitting a binary information block of $K$ bits, the $K$ most reliable polarized channels $W_{N_0}^{(i)}$ with indices $i \in \mathcal{A}$ are selected to carry these information bits; and the others with indices $i \in \mathcal{A}^{c}$ are called frozen channels and are used to transmit a fix sequence, where $\mathcal{A}^{c}$ is the complementary set of $\mathcal{A}$.
The index set $\mathcal{A}$ and its complement set $\mathcal{A}^{c}$ are called the information set and the frozen set, respectively.

\begin{figure}[!t]
  \centering
  \includegraphics[width=0.9\columnwidth]{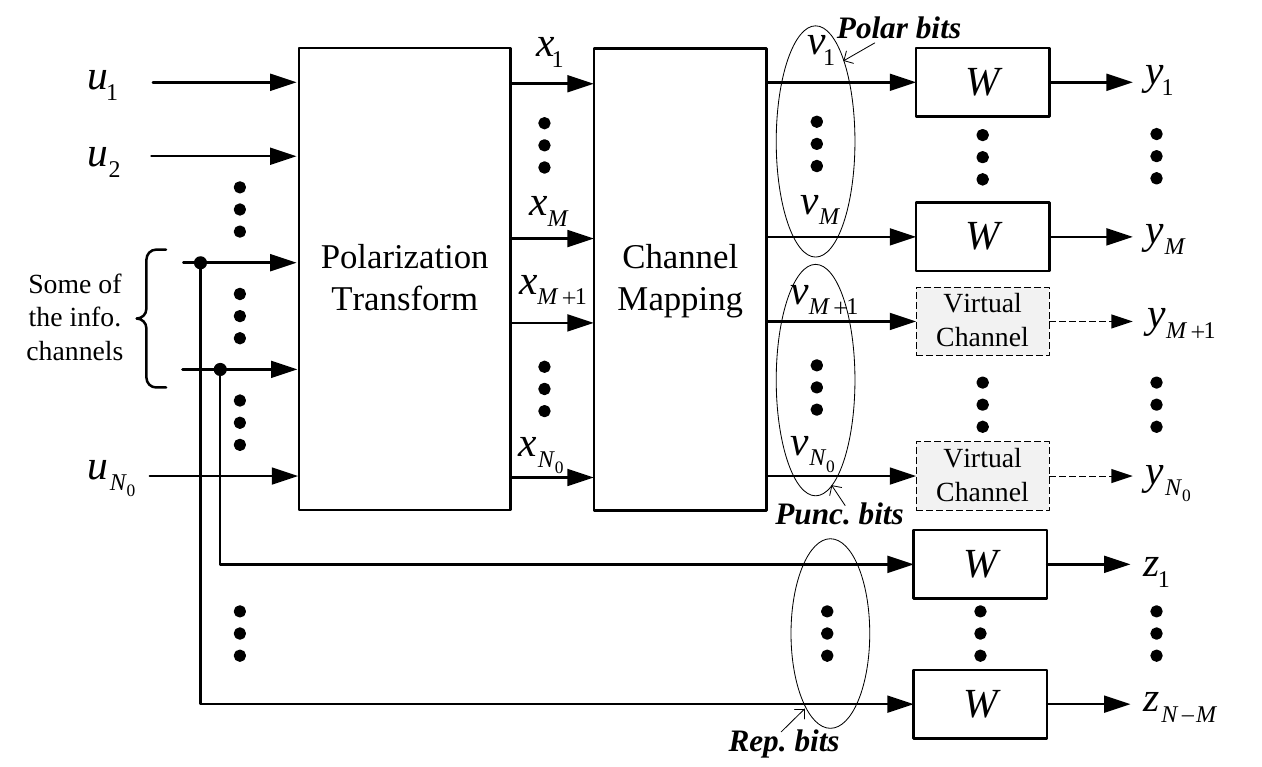}
  \caption{Channel transform of RCP coding.}
  \label{fig_chmodel}
\end{figure}

Different from the conventional polar codes, ${N_0}-M$ output bits of the polar encoder should be punctured when constructing a punctured polar code.
Therefore, before performing the polarization transform, the underlying channel uses corresponding to these punctured positions should be replaced by virtual channels which have the same input and output alphabets as $W$ but with zero capacities \cite{Niu_icc}.
As the performance of the punctured polar codes relies heavily on the specific puncturing pattern, an efficient near-optimal puncturing scheme is provided in \cite{Niu_icc}.

After the punctured polar code is constructed,  some error-prone information bits are duplicated and collected into a sequence of $N-M$ repetition bits.
The repetition bits are directly sent into another $N-M$ uses of $W$, these channel uses are called the \emph{repetition channels}.
The channel transform of the RCP coding is shown in Fig.\ref{fig_chmodel}.
With a slight abuse of notation, we write $W_N^{(i)}$ with $i=1,2,\cdots,N_0$ to denote the resulting (information and frozen) channels, where the subscript $N$ denotes the number of actual channel uses and the superscript $i$ denotes the channel index.
Note that $N$ and $N_0$ do \emph{not} have to take the same value under this setting.

When constructing a RCP code, the mappings between the repetition channels and the information channels are determined one by one. Each time, one of the undetermined repetition channels and the least reliable information channel are tied together.
Thus, the reliability of this information channel is improved and updated.
This procedure continues until all the repetition channels are determined.
The mappings between the information channels and the repetition channels are recorded in a $(N-M)$-dimensional vector $\textbf{r}$, where the \mbox{$k$-th} repetition channel carries the same bit with the \mbox{${r}_k$-th} information channel, $k=1,2,\cdots, N-M$.
The detail description is summarized in \mbox{Algorithm \ref{alg_repseq}}, where the operation $\varoast$ denotes the convolution of two LLR pdfs.
\begin{algorithm}[!ht]
\caption{Determine Repetition Indices }
\renewcommand{\algorithmicrequire}{\textbf{Input:}}
\renewcommand{\algorithmicensure}{\textbf{Output:}}
\label{alg_repseq}
\begin{algorithmic}[1]
\REQUIRE Code length of the RCP code $N$;\\
\quad \:\:Code length of the punctured polar code $M$;\\
\quad \:\:LLR pdf $\textbf{a}_{N_0}^{(i)}$ of $W_{N_0}^{(i)}$ for all $i\in \mathcal{A}$ ; \\
\quad \:\:LLR pdf $\textbf{a}$ of $W$ when $0$ is transmitted;
\ENSURE Repetition vector \textbf{r};\\
\STATE For every $W_{N}^{(i)}$, initialize its LLR pdf as $\textbf{a}_{N}^{(i)} \gets \textbf{a}_{N_0}^{(i)}$;
\STATE Calculate $P_e(W_{N}^{(i)})$ using (\ref{equ_bitpe});
\FOR{$k \gets 1 : N-M$}
    \STATE The index $i \in \mathcal{A}$ with the largest $P_e(W_{N}^{(i)})$ is found;
    \STATE Determine the repetition mapping, ${r}_k \gets i$;
    \STATE Update the LLR pdf of $W_{N}^{(i)}$ with 
    \begin{equation}
    \label{equ_conv_var_inalg}
    \textbf{a}_{N}^{(i)} \gets \textbf{a}_{N}^{(i)} \varoast \textbf{a}
    \end{equation}
    \STATE Re-calculate $P_e(W_{N}^{(i)})$ using (\ref{equ_bitpe});
\ENDFOR
\RETURN $\textbf{r}$;
\end{algorithmic}
\end{algorithm}

Since the dependencies among the information channels are not affected by adding the repetition channels, RCP codes can be decoded using a successive cancellation (SC) scheme in the same way with the conventional polar codes \cite{Arikan_first}. The block error rate (BLER) of an $(N, K, M)$ RCP code under SC decoding can be evaluated as
\begin{equation}
\label{equ_bler}
P_{B}(N, K, M) = \sum \nolimits _{i\in\mathcal{A}}{P_e(W_{N}^{(i)})}
\end{equation}

\section{HARQ Schemes based on RCP codes}
\label{section_harq}

The code length of RCP codes can be easily adjusted by adding or reducing the repetition bits.
Hence, the design of the HARQ scheme in Fig.\ref{fig_diagram} is to construct a set of $(N_{t}, K, M)$ RCP codes, where $t=1,2,\cdots, T$.
This section first gives an approximation of the throughput efficiency over BAWGNC, then the construction algorithm of the proposed \mbox{HARQ} scheme is described in detail.

\subsection{An Approximate Bound of Throughput Efficiency}
\label{section_harq:bound}
An HARQ scheme based on a set of $(N_{t}, K, M)$ RCP codes, where $t=1,2,\cdots, T$, is used to transmit an information block of $K$ bits.
After $t$ transmissions, a total of $N_t$ bits are received from the channel.
Let $E_t$ with $t=1,2,\cdots, T$ denote the event that the information block \emph{cannot} be correctly decoded using all the $N_t$ bits received in the first $t$ transmissions, and $\overline{E_t}$ denote the complementary event of $E_t$.
We write $\Pr({E_t})$ to denote the probability of event $E_t$.
Particularly, we write $E_0$ to denote the event that the information block is not known at the receiver before transmitting any bits. Obviously, $\Pr(E_0)=1$.

In transmitting information blocks of $K$-bits, the average numbers of the successfully received information bits $E[K]$ and the total transmitted bits $E[N]$ are as follows:
\begin{eqnarray}
E\left[ K \right]
\label{equ_EK}
&=&K\cdot \left( 1-\Pr\left( {E_T} \cap {E_{T-1}} \cdots \cap {E_{0}} \right) \right)
\end{eqnarray}
\begin{eqnarray}
E\left[N\right] &=& \sum \limits _{t=1}^{T}{N_t \cdot \Pr( \overline{E_t} \cap {E_{t-1}} \cap {E_{t-2}} \cdots \cap {E_{0}})}\nonumber \\
\label{equ_EN}
                &+& N_T \cdot \Pr( {E_T} \cap {E_{T-1}} \cdots \cap {E_{0}})
\end{eqnarray}

Then, the throughput efficiency can be written as
\begin{equation}
\label{equ_throughput}
\eta = \frac{E[K]}{E[N]}
\end{equation}

Obviously, we have
\begin{eqnarray}
\label{equ_a}
\Pr( {E_t} \cap  \cdots \cap {E_{0}}) \le \Pr(E_t)
\end{eqnarray}
\begin{eqnarray}
\label{equ_b}
\Pr( \overline{E_t} \cap \cdots \cap {E_{0}})=\Pr( {E_{t-1}} \cap \cdots \cap {E_{0}}) \qquad \nonumber \\
                    -\Pr( {E_{t}} \cap \cdots \cap {E_{0}})
\end{eqnarray}

However, (\ref{equ_b}) is still difficult to evaluate.
In this paper, we would like to use the following approximation
\begin{eqnarray}
\label{equ_hypothesis}
\Pr \left( {{E}_{t-1}}\cap \cdots \cap {{E}_{0}} \right)-\Pr \left( {{E}_{t}}\cap \cdots \cap {{E}_{0}} \right) \qquad \nonumber\\
\approx \Pr({{E}_{t-1}})-\Pr({{E}_{t}})
\end{eqnarray}

\begin{figure}[!t]
  \centering
  \includegraphics[width=0.8\columnwidth]{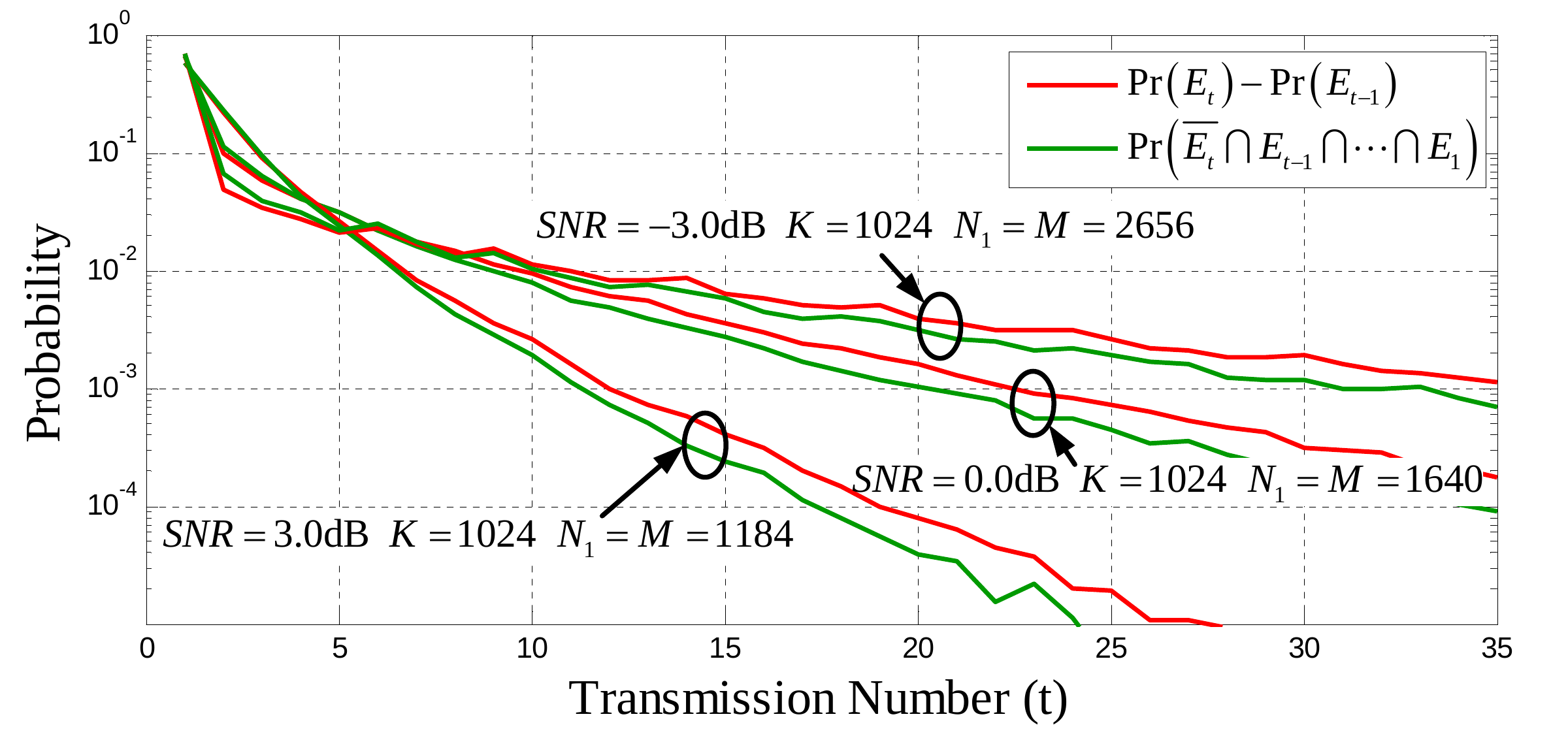}
  \caption{$\Pr({{E}_{t-1}})-\Pr({{E}_{t}})$ is used to approximate $\Pr( \overline{E_t} \cap \cdots \cap {E_{0}})$.}
  \label{fig_bound}
\end{figure}

\mbox{Fig. \ref{fig_bound}} gives the curves of $\Pr({{E}_{t-1}})-\Pr({{E}_{t}})$ and $\Pr( \overline{E_t} \cap \cdots \cap {E_{0}})$ under three transmission schemes with symbol signal-to-noise ratio (SNR) of $\{-3.0, 0, 3.0\}$dB.
In each retransmission, $30$ repetition bits are transmitted, i.e., $N_{t}-N_{t-1}=30$, where $t \ge 2$, and each curve is obtained by simulating at least $10^6$ information blocks.
As shown in \mbox{Fig. \ref{fig_bound}}, $\Pr({{E}_{t-1}})-\Pr({{E}_{t}})$ tends to be an upperbound of $\Pr( \overline{E_t} \cap \cdots \cap {E_{1}})$.

\begin{figure}[!t]
  \centering
  \includegraphics[width=0.8\columnwidth]{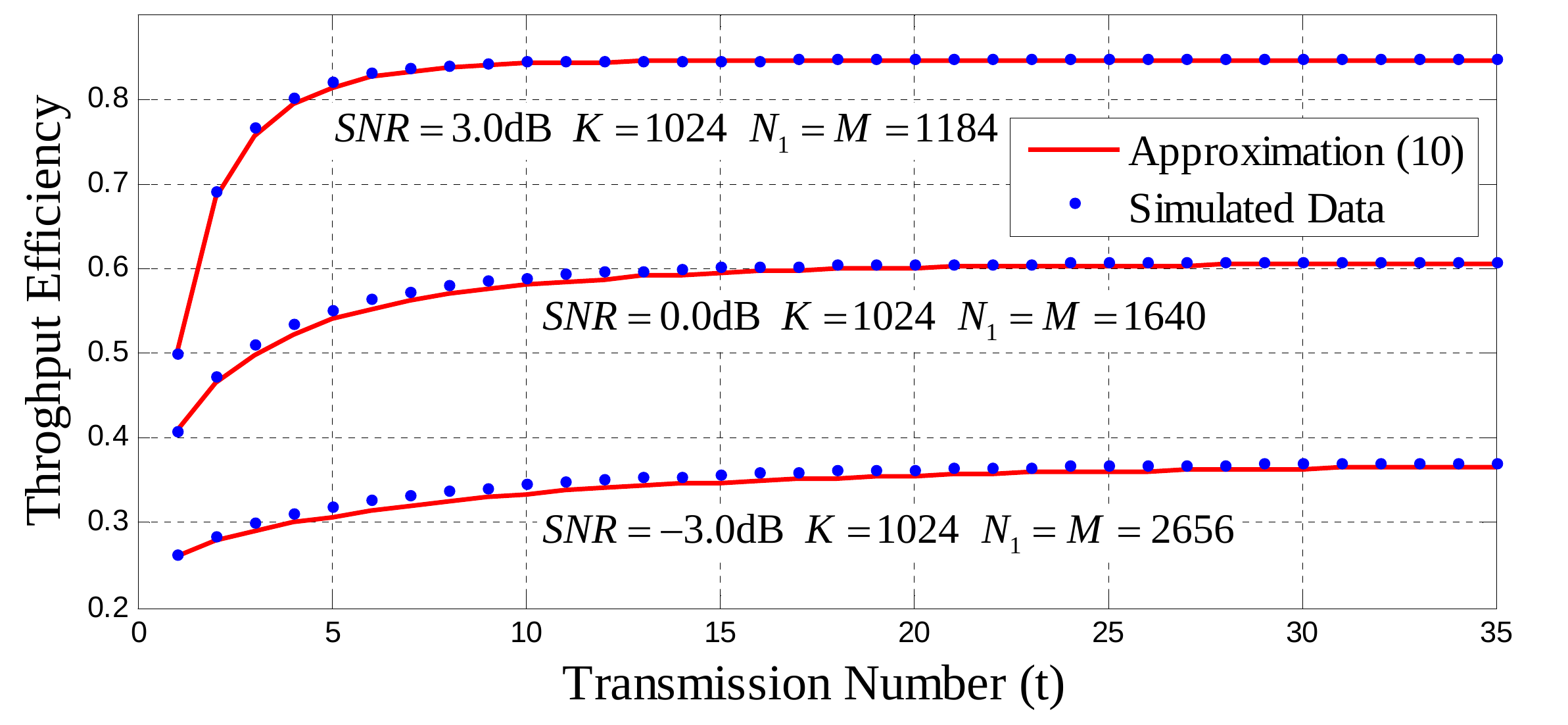}
  \caption{The approximation (\ref{equ_eta}) is usually a lowerbound.}
  \label{fig_bound_thput}
\end{figure}

Therefore, the throughput in (\ref{equ_throughput}) can be approximately calculated as
\begin{eqnarray}
\label{equ_eta}
\eta \approx \frac{K \cdot (1-\Pr(E_T))}{ \sum \nolimits_{t=1}^{T}{ N_t \cdot (\Pr(E_{t-1})-\Pr(E_t)) } + N_T \cdot \Pr(E_T) }
\end{eqnarray}
where $\Pr({E_t})$ is in fact the BLER of the $(N_{t}, K, M)$ RCP code and can be evaluated by (\ref{equ_bler}).

Moreover, since both the substitutions of $\Pr(E_t)$ for $\Pr( {E_t} \cap  \cdots \cap {E_{0}})$ in (\ref{equ_a}) and $\Pr(E_{t-1})-\Pr(E_t)$ for $\Pr( \overline{E_t} \cap {E_{t-1}} \cap \cdots \cap {E_{1}})$ in (\ref{equ_hypothesis}) are upperbounds, the approximation (\ref{equ_eta}) tends to be a lowerbound of $\eta$ (as shown in Fig. \ref{fig_bound_thput}).

\subsection{Design An HARQ Scheme}
\label{section_harq:design}
Utilizing (\ref{equ_eta}), an HARQ scheme with information block size $K$ can be designed via a greedy search described in \mbox{Algorithm \ref{alg_harq}}, where the function $\textsl{sort}(\mathcal{X})$ puts the elements of $\mathcal{X}$ into a vector $\textbf{x}$ in ascending order, i.e., for all $1\le i\le j\le |\mathcal{X}|$, $x_i \in \mathcal{X}$ and $x_i \le x_j$.
The algorithm outputs a vector $\textbf{s}$ where its elements means that, the RCP codes are constructed with $M=s_1$, and after the \mbox{$t$-th} transmission attempt, totally $s_{t+1}$ bits should be transmitted, i.e., $N_t=s_{t+1}$.

\begin{algorithm}[!ht]
\caption{Design an HARQ Scheme Based on RCP codes}
\renewcommand{\algorithmicrequire}{\textbf{Input:}}
\renewcommand{\algorithmicensure}{\textbf{Output:}}
\label{alg_harq}
\begin{algorithmic}[1]
\REQUIRE Information block length $K$;\\
\quad \: Maximum number of transmission times $T$; \\
\quad \: Maximum transmitted bits $Q$;
\ENSURE Transmission scheme vector \textbf{s};
\STATE Initialize $\mathcal{S}_{opt} \gets \phi$, $\eta_{opt} \gets 0$, $m_{opt} \gets 0$;
\FOR{$m \gets K:Q$}
    \STATE $//$ Compute the BLERs with all possible code length.
    \STATE Allocate a $(Q-m+1)$-dimensional vector $\textbf{e}$;
    \FOR{$n \gets m:Q$}
    \STATE Construct an $(n, K, m)$ RCP codes;
    \STATE ${e}_{n-m+1} \gets P_B(n, K, m)$;
    \ENDFOR
    \STATE $//$ Search for the optimal scheme with $M=m$.
    \STATE $\mathcal{S} \gets \phi$, $\eta \gets 0$;
    \FOR{$l \gets 1:T$}
        \STATE Allocate a temporary set $\mathcal{T} \gets \mathcal{S}$;
        \FOR{$n \gets m:Q$ and $n \notin \mathcal{S}$}
            \STATE $\textbf{t} \gets \textsl{sort}(\mathcal{S}\cup \{n\})$;
            \STATE Calculate the throughput of the temporary scheme:
            \begin{eqnarray}
            \!\!\!\!\!\!\lambda &\gets& \sum \nolimits_{i=1}^{l}{ t_i \cdot (e_{t_{i-1}-m+1}-e_{t_i-m+1}) } \\
            \label{equ_eta_inalg}
            \!\!\!\!\!\!\rho &\gets& \frac{K \cdot (1-e_{t_l-m+1})}{\lambda + t_l \cdot e_{t_l-m+1}}
            \end{eqnarray}
            \IF{$\rho > \eta$}
                \STATE $\mathcal{T} \gets \mathcal{S} \cup \{n\}$, $\eta \gets \rho$;
            \ENDIF
        \ENDFOR
    \STATE Record the locally best scheme, $\mathcal{S} \gets \mathcal{T}$;
    \ENDFOR
    \IF{$\eta > \eta_{opt}$}
        \STATE \mbox{$\mathcal{S}_{opt} \gets \mathcal{S}$}, \mbox{$\eta_{opt} \gets \gamma$}, \mbox{$m_{opt} \gets m$};
    \ENDIF
\ENDFOR
\STATE Allocate a $(T+1)$-dimensional vector $\textbf{s}$;
\STATE $s_1 \gets m_{opt}$, $\textbf{s}_{2:T+1} \gets \textsl{sort}(\mathcal{S}_{opt})$;
\RETURN $\textbf{s}$;
\end{algorithmic}
\end{algorithm}

In \mbox{Algorithm \ref{alg_harq}}, the outer loop at line 2 is executed $Q-K+1$ times.
For each $m$ in the loop, the operations in line $5$ to $8$ are equivalent to construct a punctured polar code with code length $m$ at first, and then construct a set of \mbox{RCP} codes by adding the repetition channels one by one.
Each time, only one additional convolution defined in (\ref{equ_conv_var_inalg}) is required.
Taking the outer loop into account, a total of $\sum \nolimits _{m=K}^{Q}{(\overline{m} \log \overline{m}+Q-m)}$ convolutions are required, where $\overline{m}=2^{\lceil \log m \rceil}$.
Besides, the loops in line $11$ to $21$ take time complexity $O(T\cdot(Q-m))$.
Since $T$ is usually much smaller than $K$ and $Q$, the operations in the loop between line $5$ and $8$ dominate the complexity of the whole algorithm.
Therefore, the complexity of \mbox{Algorithm \ref{alg_harq}} is upper bounded by $O(Q^2 \log Q)$.

\section{Simulation Results}
\label{section_simulation}
In this section, the performance of the proposed HARQ scheme based on RCP codes is evaluated via simulations over binary-input additive white Gaussian noise channels (BAWGNCs).
The RCP codes are with information block lengths $K=1024$ and decoded using SC decoding algorithm.
For efficient construction, Gaussian approximation (GA) for DE \cite{Trifonov_GA} is used to evaluate the error probabilities in (\ref{equ_bitpe}).

\begin{figure}[!t]
  \centering
  \includegraphics[width=0.8\columnwidth]{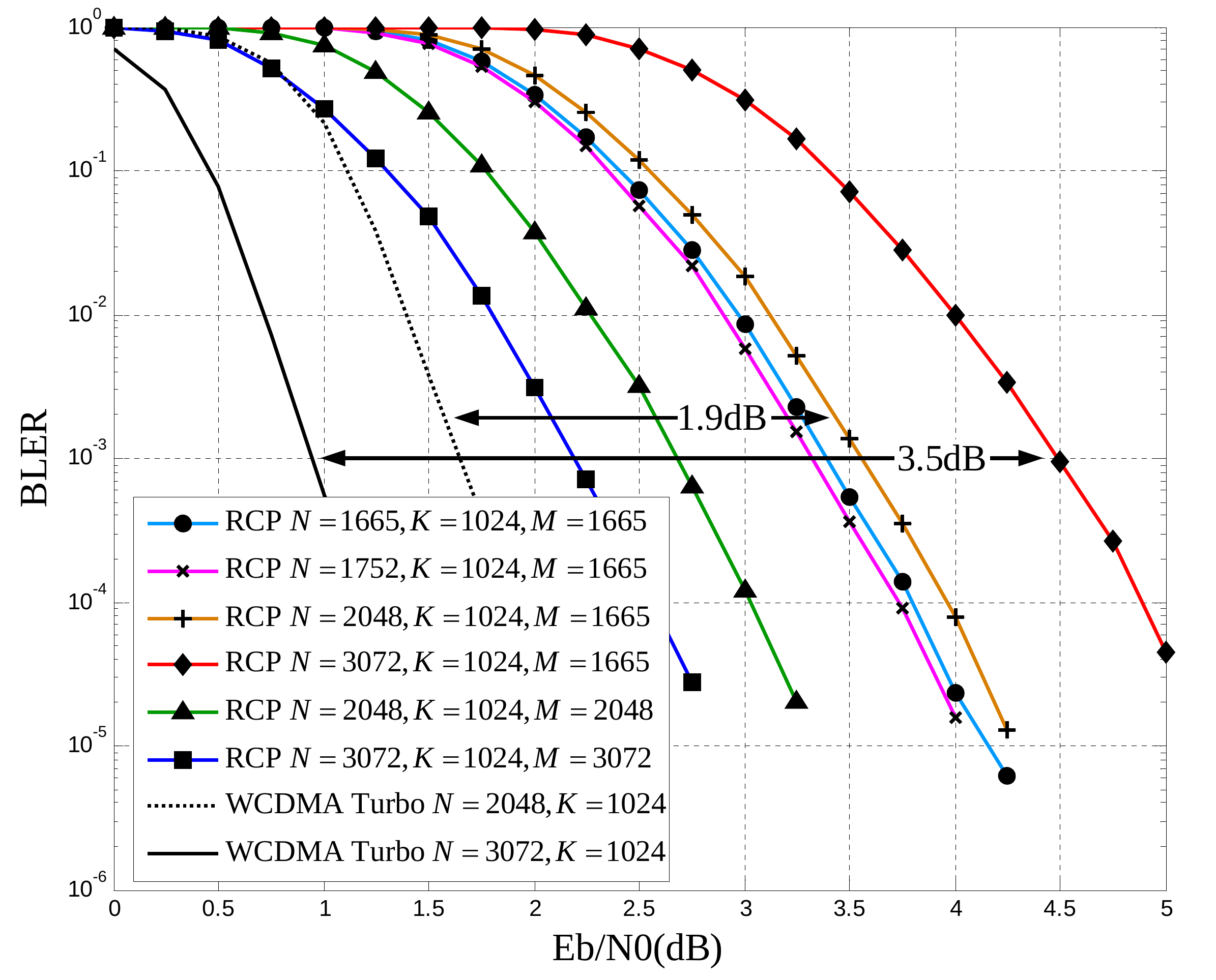}
  \caption{Performance of RCP codes over BAWGN channels.}
  \label{fig_rcprp}
\end{figure}

Intuitively, RCP codes are not good codes.
This can be confirmed by the simulation results over BAWGNCs shown in Fig. \ref{fig_rcprp}.
In comparison, the BLER performance of the turbo code used in WCDMA systems \cite{3GPP_25212} is also given.
We can see that, there is a gap of $2.0-3.5$dB between the simulated RCP codes and the turbo codes at the BLER around $1.0\times10^{-3}$, and the longer the repetition sequence is the worse the performance will be.
An extreme case is that when the code length $N$ goes much larger than $M$, the codes look more like the naive repeat codes which do not have any coding gain.

\begin{figure}[!t]
  \centering
  \includegraphics[width=0.8\columnwidth]{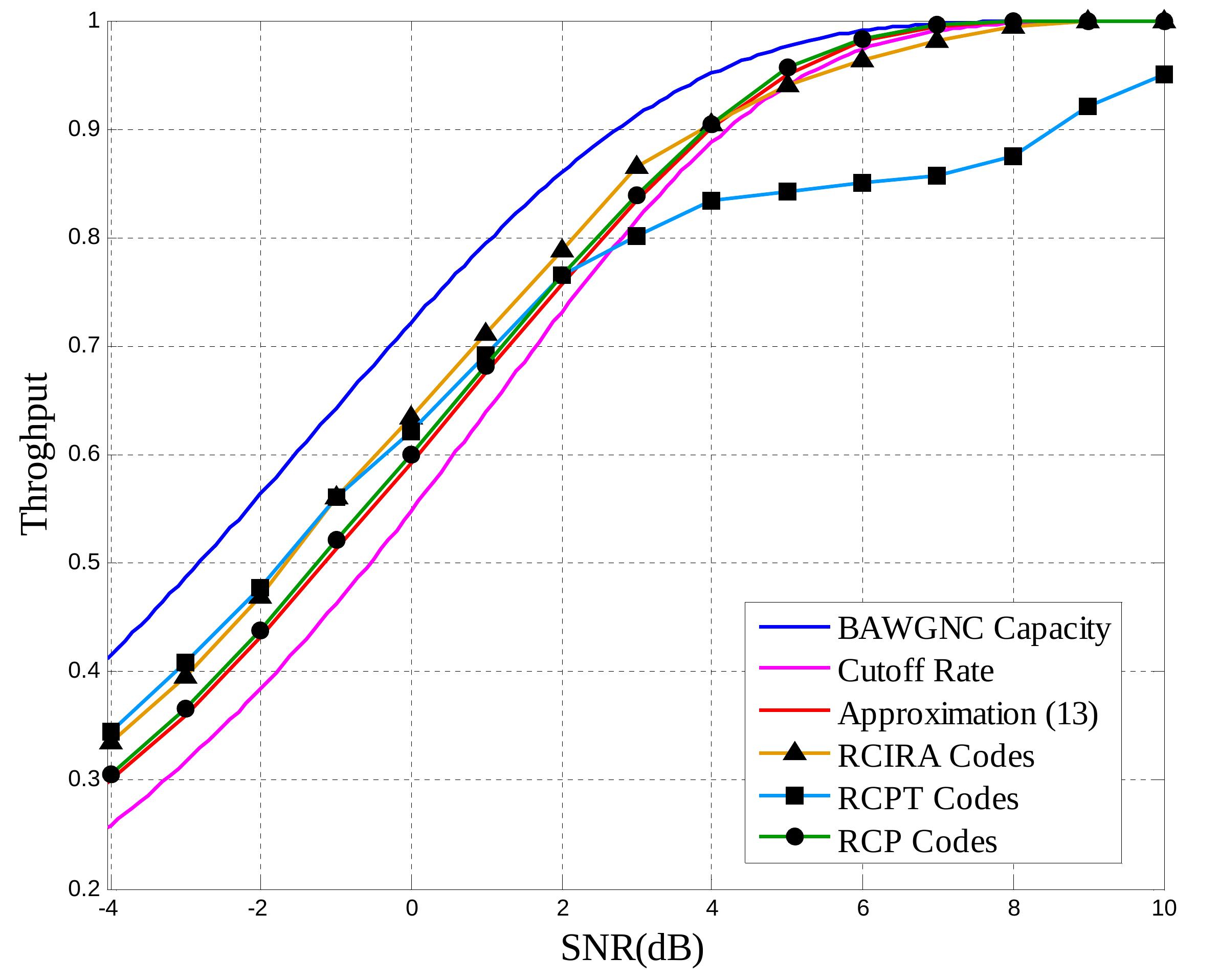}
  \caption{Throughput of HARQ transmission scheme employing RCP codes with $K=1024$ in BAWGN channels, where results for RCIRA codes are from \cite{Yue_RCIRA} ($K=512$) and RCPT codes are from \cite{Rowitch_RCPT} ($K=1024$).}
  \label{fig_sim_awgn}
\end{figure}

However, when applying the RCP codes to HARQ schemes, the results are quite interesting.
Fig. \ref{fig_sim_awgn} shows the throughput efficiency of the proposed HARQ scheme over \mbox{BAWGNCs}.
In comparison, the throughput efficiency curves of the \mbox{HARQ} schemes based on the rate-compatible punctured turbo codes (\mbox{RCPT}) \cite{Rowitch_RCPT} and the rate-compatible irregular repeat-accumulate (\mbox{RCIRA}) codes \cite{Yue_RCIRA} (as a representative LDPC codes) are provided.
As the figure shows, the RCP coded schemes can work as well as those with turbo codes or LDPC codes and are only about $1.0-1.5$dB from the capacity.
In the low SNR regime, RCP coded schemes are a little worse than those of turbo and LDPC, the performance loss are as small as only about $0.5$dB.
As the SNR goes higher, the performance gap between RCP and turbo/LDPC coded schemes becomes smaller.
When the SNR is above $4.0$dB, the proposed scheme achieves better throughput efficiency than that of turbo/LDPC.

\section{Conclusions}
\label{section_conclusions}

An HARQ scheme based on RCP codes is proposed.
Simulation results over BAWGNCs show that, the proposed transmission scheme under the SC decoding can perform as well as the existing schemes based on turbo codes and LDPC codes, and are only about 1.0-1.5dB away from the channel capacity with information block length of 1024 bits.


%

\ifCLASSOPTIONcaptionsoff
  \newpage
\fi

\end{document}